\documentclass[runningheads]{llncs} \pdfoutput=1 \usepackage{graphicx}
 \usepackage{amssymb}

\begin{document}

\makeatletter \def\imod#1{\allowbreak\mkern10mu({\operator@font
    mod}\,\,#1)} \makeatother

\newcommand{\eprint}[1]{{\tt #1}} \newcommand{\eqn}[1]{(\ref{eq.#1})}
\newcommand{\bra}[1]{\mbox{$\left\langle #1 \right|$}}
\newcommand{\ket}[1]{\mbox{$\left| #1 \right\rangle$}}
\newcommand{\braket}[2]{\mbox{$\left\langle #1 | #2 \right\rangle$}}
\newcommand{\av}[1]{\mbox{$\left| #1 \right|$}}
\newcommand{\ve}{\varepsilon} \newcommand{\osc}{{\mbox{\rm \scriptsize
      osc}}} \newcommand{\tot}{{\mbox{\rm \scriptsize tot}}}
\newcommand{\lga}{{\mbox{\rm \scriptsize LGA}}}
\newcommand{\swap}{{\mbox{\rm \scriptsize swap}}}
\newcommand{\ground}{{\mbox{\rm \scriptsize ground}}}
\newcommand{\cycle}{{\mbox{\rm \scriptsize cycle}}}
\newcommand{\particle}{{\mbox{\rm \scriptsize particle}}}
\newcommand{\internal}{{\mbox{\rm \scriptsize internal}}}
\newcommand{\nonrel}{{\mbox{\rm \scriptsize non-rel}}}
\newcommand{\twoparticles}{{\mbox{\rm \scriptsize 2p}}}
\newcommand{\block}{{\mbox{\rm \scriptsize block}}}
\newcommand{\blockchange}{{\mbox{\rm \scriptsize block-change}}}
\newcommand{\statechange}{{\mbox{\rm \scriptsize state-change}}}
\newcommand{\even}{{\mbox{\rm \scriptsize even}}}
\newcommand{\odd}{{\mbox{\rm \scriptsize odd}}}
\newcommand{\rt}{{\mbox{\rm \scriptsize R}}}
\newcommand{\lt}{{\mbox{\rm \scriptsize L}}}
\newcommand{\ball}{{\mbox{\rm \scriptsize ball}}}
\newcommand{\shift}{{\mbox{\rm \scriptsize shift}}}
\newcommand{\D}[2]{\frac{\partial #2}{\partial #1}}

\newcommand{\sinc}{{\mbox{\rm sinc}}\:}
\newcommand{\sincs}{{\mbox{\rm sinc$^2$}}}

\newcommand{\mom}{{\mbox{\rm \scriptsize mom}}}
\newcommand{\motion}{{\mbox{\rm \scriptsize motion}}}
\newcommand{\smax}{{\mbox{\rm \scriptsize max}}}
\newcommand{\smin}{{\mbox{\rm \scriptsize min}}}

\newcommand{\up}{{\texttt{\tiny $\setminus/$}}}
\newcommand{\dn}{{\texttt{\tiny $/\setminus$}}}

\newcommand{\p}{{\texttt{\tiny $+$}}}
\newcommand{\m}{{\texttt{\tiny $-$}}}

\setlength{\fboxsep}{.1pt} \setlength{\fboxrule}{.1pt}


\title{Finite-State Classical Mechanics}


 \author{Norman Margolus}

\institute{Massachusetts Institute of Technology, Cambridge MA, USA.
  \email{nhm@mit.edu}}

\maketitle

\begin{abstract}
Reversible lattice dynamics embody basic features of physics that
govern the time evolution of classical information.  They have finite
resolution in space and time, don't allow information to be erased, and
easily accommodate other structural properties of microscopic physics,
such as finite distinct state and locality of interaction.  In an ideal
quantum realization of a reversible lattice dynamics, finite classical
rates of state-change at lattice sites determine average energies and
momenta.  This is very different than traditional continuous models of
classical dynamics, where the number of distinct states is infinite,
the rate of change between distinct states is infinite, and energies
and momenta are not tied to rates of distinct state change. Here we
discuss a family of classical mechanical models that have the
informational and energetic realism of reversible lattice dynamics,
while retaining the continuity and mathematical framework of classical
mechanics.  These models may help to clarify the informational
foundations of mechanics.
\end{abstract}

\section{Introduction}\label{sec.intro}

The physics of continuous classical materials and fields is
pathological.  For example, in classical statistical mechanics, each
degree of freedom of a system at thermal equilibrium has about the same
finite average energy, proportional to the temperature.  This implies
that a continuous material---which has an infinite number of degrees of
freedom---will be able to absorb energy at a finite rate forever
without heating up.  Exactly this pathology, evident in the radiation
field inside hot cavities (black bodies), led to the overturn of
classical mechanics as a fundamental theory and the advent of quantum
mechanics \cite{planck}.

A similar pathology of the continuum exists in the classical mechanics
of particles.  Quantum mechanics provides a fundamental definition of
energy in terms of rate of change in time: {\em frequency}.  What we
call energy in the classical realm is (in fundamental units) just the
average frequency of a large quantum system---and this is precisely the
maximum rate at which the system can transition between perfectly
distinct states \cite{max-speed,counting,max-avg}.  Since all physical
systems have finite energy, they all have a finite rate of distinct
state change.  But in continuous classical mechanics, each
infinitesimal interval of time brings a perfectly distinct new state,
and so the rate of state change is infinite.  Similarly, finite
momentum only allows a finite rate of distinct state change due to
motion, not the infinite rate required by infinite resolution in space.

It would be nice to have a version of classical dynamics that avoids
the pathologies of infinite state and infinite resolution in space and
time, while still being a subset of ordinary classical mechanics.
Fredkin's billiard ball model of computation \cite{bbm} illustrates
that this is possible: a carefully designed classical mechanical system
with discrete constraints on initial conditions can be equivalent, at
discrete times, to a reversible finite-state dynamics.  In this paper,
we discuss Fredkin's model as well as others where the equivalence is
even more direct.  In these examples, local rates of state change in
the finite-state dynamics play the roles of mechanical energies and
momenta.  This property is physically realistic, and arises from the
fact that ordinary reversible computations, such as these, can be
interpreted as special cases of quantum computations
\cite{bennett-review}, and hence inherit quantum definitions of energy
and momentum based on rates of state change.

\section{Energy and momentum of reversible lattice gases}

Some of the simplest models of physical systems are lattice models with
classical finite state. Dynamical models of this sort with local
interactions are often referred to as {\em cellular automata}, but here
we will favor the more physical term {\em classical lattice gas}, which
encompasses both deterministic and stochastic physical models
\cite{ruelle,stanley,cam-book,nks,rothman,chopard,rivet}.  We discuss
reversible classical lattice gases as foundational models of both
classical and quantum mechanics.

By foundational models we mean here the simplest examples of systems
that exactly incorporate basic physical properties and principles ({\em
  cf.}  \cite{toffoli-pde}). That a world with finite entropy has
classical lattice gas foundations is not surprising, and is well
accepted in statistical mechanics \cite[\S 2.4]{ruelle}, where
finite-state lattice models have provided great insight into the
foundations of the field.  The idea that classical lattice gases are
foundational is, however, much less accepted in ordinary mechanics,
where the close relationship of reversible finite-state lattice gases
to continuous-time and continuous-space mechanics---and the fundamental
link that energy and momentum provide---are not widely appreciated.

\subsection{Continuous space and time}

The fundamental models of mechanics are continuous in space and time,
and hence seem very non-discrete.  In fact, though, all realistic
physical systems have finite resolution in space and time, usually
described using continuous mathematics, and similar descriptions can be
applied to discrete models.

The finite resolution in space and time of quantum systems can be
expressed in terms of uncertainty relations \cite{heisenberg}, but is
better thought of as akin to the effective discreteness of a
finite-bandwidth classical signal \cite{counting,max-avg}.
Interestingly, this kind of discreteness was recognized around the same
time that the founders of quantum mechanics discovered uncertainty
\cite{nyquist}.  The discoverer, Harry Nyquist, was thinking about how
many dots and dashes could be put into a telegraph signal, and he
realized that bandwidth was the key quantity that set the bound.

He gave a simple argument, first considering a signal periodic in time,
and then generalizing to the average rate for an infinitely long
period. Consider a complex valued periodic wave.  This is composed of a
discrete set of Fourier components that fit the period: for period $T$
the possible frequencies are $1/T$, $2/T$, $3/T$, etc.  With a limited
range of frequencies (a limited {\em bandwidth}), the Fourier sum
describing the wave has only a finite number of terms with a finite
number (say $N$) coefficients.  With $N$ coefficients we can only
freely choose the value of the sum at $N$ times.  Thus the minimum
range of frequencies $\nu_\smax-\nu_\smin$ needed to have $N$ distinct
values of the sum is given by the minimum separation between
frequencies $1/T$, times the minimum number of separations $N-1$:
\begin{equation}\label{eq.wt}
  \nu_\smax-\nu_\smin \ge \frac{N-1}{T}\;.
\end{equation}
For a long wave, the bandwidth $\nu_\smax-\nu_\smin$ is the maximum
average density of distinct values, $N/T$.  Turning the argument
around, if we know the values of a periodic signal with finite
bandwidth at enough discrete points, we can determine all coefficients
in the finite Fourier sum: the rest of the continuous wave is then
determined and carries no additional information
\cite{interpolation}. Thus waves with finite bandwidth are effectively
discrete.

\subsection{Finite resolution in quantum mechanics}

The argument above goes through essentially unchanged for the wavelike
evolution of quantum systems.  The wavefunction for an isolated system
is expressed in the energy basis as a superposition of frequency
components: $\nu_n=E_n/h$, where $E_n$ is the $n$-th energy eigenvalue,
and $h$ is Planck's constant.  To have $N$ distinct (mutually
orthogonal) states in a periodic time evolution, there must be a
superposition of at least $N$ distinct energy eigenfunctions, with
distinct frequencies.  Again the minimum frequency separation is $1/T$
if the period is $T$, so the minimum range of energy eigenfrequencies
is again given by \eqn{wt}.

For systems that exactly achieve this bound on orthogonal evolution, it
is easy to show that the $N$ equally spaced frequency components must
be equally weighted, and that the same minimum-bandwidth distribution
minimizes all reasonable measures of frequency width
\cite{counting,max-avg}.  For example, for the minimizing distribution,
average frequency $\bar\nu$ minus the lowest $\nu_\smin$ is half the
bandwidth, so
\begin{equation}\label{eq.et}
  2(\bar\nu-\nu_\smin) \ge \frac{N-1}{T}\;.
\end{equation}
In quantum mechanics, the average energy of an isolated system is
$E=h\bar\nu$.  The lowest (ground state) energy $E_0=h\nu_\smin$ is
like the lowest frequency used in a classical signal: it is the start
of the frequency range available for the dynamics of the isolated
system.  If the ground state energy $E_0$ is taken to be zero and $N\gg
1$, then letting $\nu_\perp=N/T$ be the average density of distinct
states in time, and choosing units with $h=2$ (so $E=2\bar\nu$),
\eqn{et} becomes
\begin{equation}\label{eq.e} 
E\ge \nu_\perp \;.
\end{equation}
Thus energy is the maximum average rate of distinct state change.  This
can also be regarded as an uncertainty relation between average
orthogonalization time $\tau=1/\nu_\perp$ and average energy width:
$(E-E_0)\tau \ge 1$.  All uncertainty relations between $\tau$ and any
other width $\Delta \nu$ of the energy-frequency distribution of the
wavefunction are similar \cite{counting,max-avg}, with the choice of
width and number of distinct states changing the bound by only a factor
of order one---a periodic oscillation between just two distinct states
is the fastest \cite{max-speed,mandelstam}.  The same kind of Fourier
analysis also applies to waves in space, rather than in time.  All such
bounds are attained simultaneously for minimum bandwidth, in which case
quantum evolution becomes equivalent to a discrete evolution on a
spacetime lattice: only the values of the wavefunction at lattice
points in space and time are distinct
\cite{counting,max-avg,emulation,kempf}.  The rest of the continuous
state is redundant.

\subsection{Motion defines momentum}

A moving particle has both extra distinct states due to its distinct
positions, and extra energy due to its motion.  In the particle's rest
frame, it has neither.  For a large ($\approx$~classical) system moving
between two events, evolving at the maximum rate of distinct state
change allowed by its energy, we can use \eqn{e} in the two frames to
count the extra distinct states due to the motion.  If the system has
velocity $v$ and magnitude of momentum $p$, and the events are
separated by time $\Delta t$ and distance $\Delta x$ in the laboratory
frame, and $\Delta t_r$ in the rest frame, with $E$ and $E_r$ the
corresponding energies, the invariant time-energy interval is
\begin{equation}\label{eq.edt}
  E\Delta t - p\Delta x = E_r\Delta t_r\;.
\end{equation}
But from \eqn{e}, $E\Delta t$ is simply the number of distinct
states seen in the laboratory frame, and $E_r\Delta t_r$ the number
seen in the rest frame.  The difference is the number of distinct
states due to the motion, which from \eqn{edt} is simply $p\Delta
x$.  Thus if $\mu_\perp$ is the average density in space of states
distinct due to the motion,
\begin{equation}\label{eq.p} 
p\ge \mu_\perp\;.
\end{equation}
If we multiply \eqn{p} by $v=\Delta x/\Delta t$, we get that $v p \ge
\Delta x \,\mu_\perp/\Delta t\equiv\nu_\motion$, the number of distinct
states per unit time due to motion.  So we also have
\begin{equation}\label{eq.mom}
  p \ge \nu_\motion/v\;.
\end{equation}
Thus motion and speed define a minimum $p$.  In a quantum realization
of a reversible lattice gas, a hop of an isolated particle from one
lattice site to another is a distinct state change, and speed is just
distance over time.  In our discussion of energy and momentum
conserving lattice gases, we will use the minimum possible momentum for
any quantum system, from \eqn{mom}, as our estimate of the momentum of
a freely moving isolated particle in an ideal realization
\cite{emulation,ideal-energy}.

\subsection{Minimum energy}

Choosing units with the speed of light $c=1$, it is always true
relativistically for a freely moving particle that $E=p/v$, so we can
compute the energy of a free particle once we know its momentum and
velocity.  From \eqn{mom}, $E\ge \nu_\motion/v^2$, so energy is
smallest when $v$ is as large as possible---we can treat $\nu_\motion$
as constant here, since it doesn't depend on the distance between
lattice sites, and we can make our particles travel faster by
increasing only the distance between lattice sites without changing the
time. Now, given a lattice gas dynamics with a set of particle
velocities related by the lattice geometry, there is a family of
equivalent evolutions that only differ in the choice of the fastest
particle speed.  Of these, the evolution with the least possible energy
has its fastest-moving particles traveling at the speed of light.  This
makes sense physically, since a system with a non-zero rest-frame
energy has a non-trivial internal dynamics---time passes in the rest
frame.  If we want to just model a logical evolution and nothing extra,
the fastest-moving particles should have no internal dynamics.

\section{Finite-state classical mechanics}

We discuss three reversible lattice gases, each with a finite-state
dynamics that reproduces discrete samples of a classical particle
evolution, and one model that samples a classical field.  In its
discrete form, the field example also turns into a reversible lattice
gas. Although these models could represent macroscopic systems with any
given energy, we're interested here in looking at intrinsic minimum
energy and momentum defined by state change on the lattice.  We analyze
this for just the last two models---the first two are introductory.
The field example is particularly interesting because energy and
momentum are bound together on the lattice, moving as a relativistic
particle with a discrete set of possible speeds. This behavior is
intimately related to a biased random walk.

\subsection{Sampled particles}

\subsubsection{Lattice gas fluids.}

\begin{figure}{ 
\begin{minipage}{\textwidth} \includegraphics[height=1.2in]{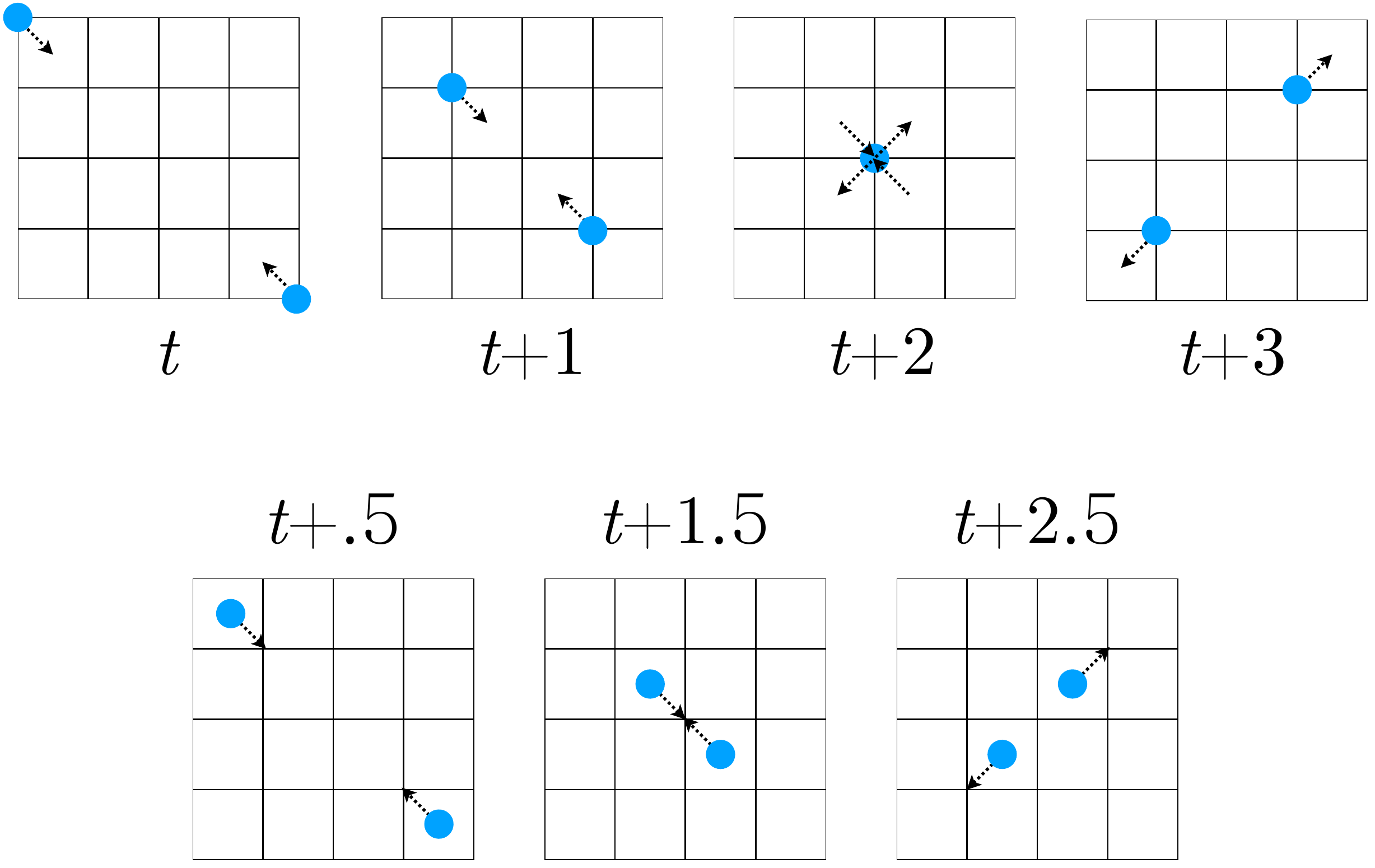} \hspace{.45in}\includegraphics[height=1.2in]{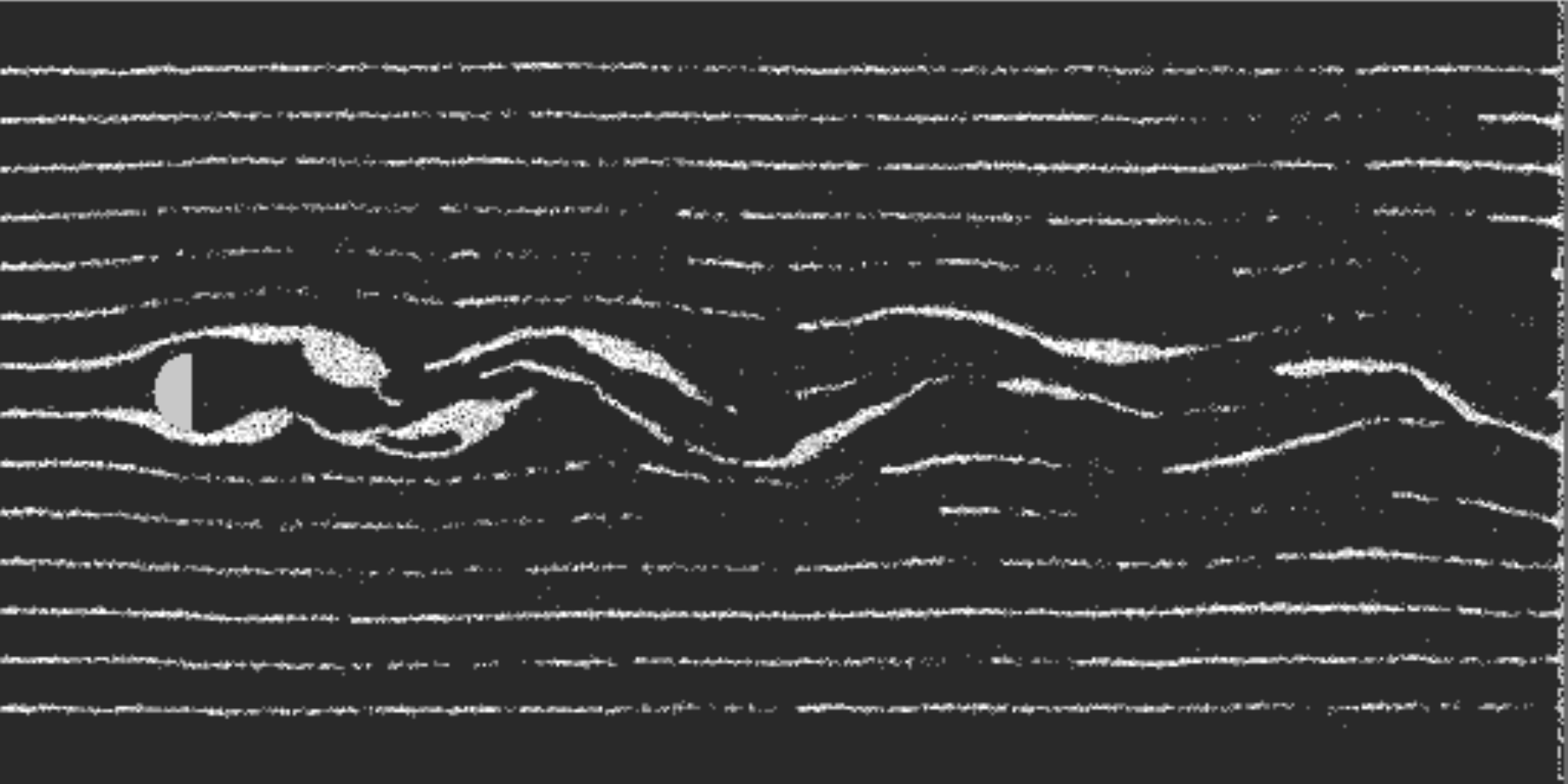} \end{minipage}}
  \caption{{\bf Lattice gas molecular dynamics.} {\em Left:} Particle
    and momentum conserving collisions in a single-speed four-velocity
    lattice gas.  Particles are at lattice points at integer times.  At
    half integer times, they are midway between.  {\em Right:} A
    single-speed six-velocity lattice gas run on two million sites of a
    triangular lattice, with obstacles and a visualizing ``smoke'' gas
    added to the model.  It exhibits realistic fluid behavior.}
\label{fig.lga}
\end{figure}

The first lattice gases that reproduced samples of a classical
mechanical evolution were models of fluids \cite{hpp}.  Lattice gas
fluids are stylized molecular dynamics models, with particles started
at points of a lattice, moving at one of a discrete set of velocities,
and colliding at points of the lattice in a manner that conserves
momentum and guarantees particles will again be on the lattice at the
next integer time.  We illustrate this in Figure~\ref{fig.lga} (Left).
We show two particles of a four-velocity 2D lattice gas.  In the top
row we show the particles at integer times, in the bottom row halfway
between integer times.  The dynamics shown is invertible and is
momentum and particle conserving, but is too stylized to be a realistic
fluid.  Four directions aren't enough to recover fully symmetric Navier
Stokes fluid flow in the large scale limit---but six are
\cite{fhp,rothman,chopard,rivet}!  Figure~\ref{fig.lga} (Right) is a
snapshot of a simulation of a six-velocity lattice gas, with obstacles
and a second tracer fluid (smoke) added, showing flow past an obstacle
\cite{super}.

The reason we can talk here about momentum conserving collisions is
because the discrete lattice gas is, conceptually, embedded in a
continuous dynamics where we know what momentum is.  In classical
mechanics, a dynamics with continuous symmetry under translations in
space defines a conserved linear momentum, and a continuous rotational
symmetry defines a conserved angular momentum \cite{noether}.  The
embedding allows us to define discrete conserved quantities that derive
from continuous symmetries which cannot exist on a discrete lattice.
The full continuous symmetries associated with the conservations
\cite{noether} can only emerge in a lattice model in the macroscopic
limit.  This makes conservation more fundamental than continuous
symmetry in lattice models \cite{dm}.

In order to emulate the reversibility of microscopic physics, a local
lattice dynamics must have a structure where data on the lattice are
partitioned into separate groups for updating \cite{ica,kari,lose};
then if the transformation of each group is invertible, this property
is inherited by the overall dynamics.  At least two different
partitions, used at different times, are required---with only one, each
group would be forever isolated.  In a continuous dynamics that has
been initialized to act discretely, the alternation of partitions does
not involve any explicit time dependence.  For example, at the
integer-times of Figure~\ref{fig.lga} (Top Left), we see that all
collisions happen at lattice locations, and the data at each lattice
site are transformed independently of all other sites---this
constitutes one partition.  Not only invertibility, but particle and
momentum conservation are guaranteed by the collision rule.  In between
collisions, particles travel straight to adjacent lattice locations
without interacting.  This constitutes the second partition.

We get a different view of partitioning for the same continuous
dynamics if we define a lattice gas from the half-integer-time states
of Figure~\ref{fig.lga} (Bottom Left).  In this case, we catch all
particles when they are going straight, in between lattice sites.
Particles are spread out in space, rather than piled up at lattice
sites, and we can tell which way they are going from where they are,
when.  Groups of four locations that can contain particles converging
on a lattice site define a partition---for example, the middle
2$\times$2 block of the middle time step.  These are replaced by their
values after the collision, independently for each block.  The outgoing
particles then converge on a new set of lattice sites, defining a
second partition---the update rule for the two partitions is the same.
From the point of view of the continuous dynamics, the $2\times2$
blocks are just imaginary boxes we've drawn around regions where
particles are converging at a given moment.


\subsubsection{The billiard ball model.}

\begin{figure}[t]{ 
\begin{minipage}{\textwidth}
  \includegraphics[height=1.2in]{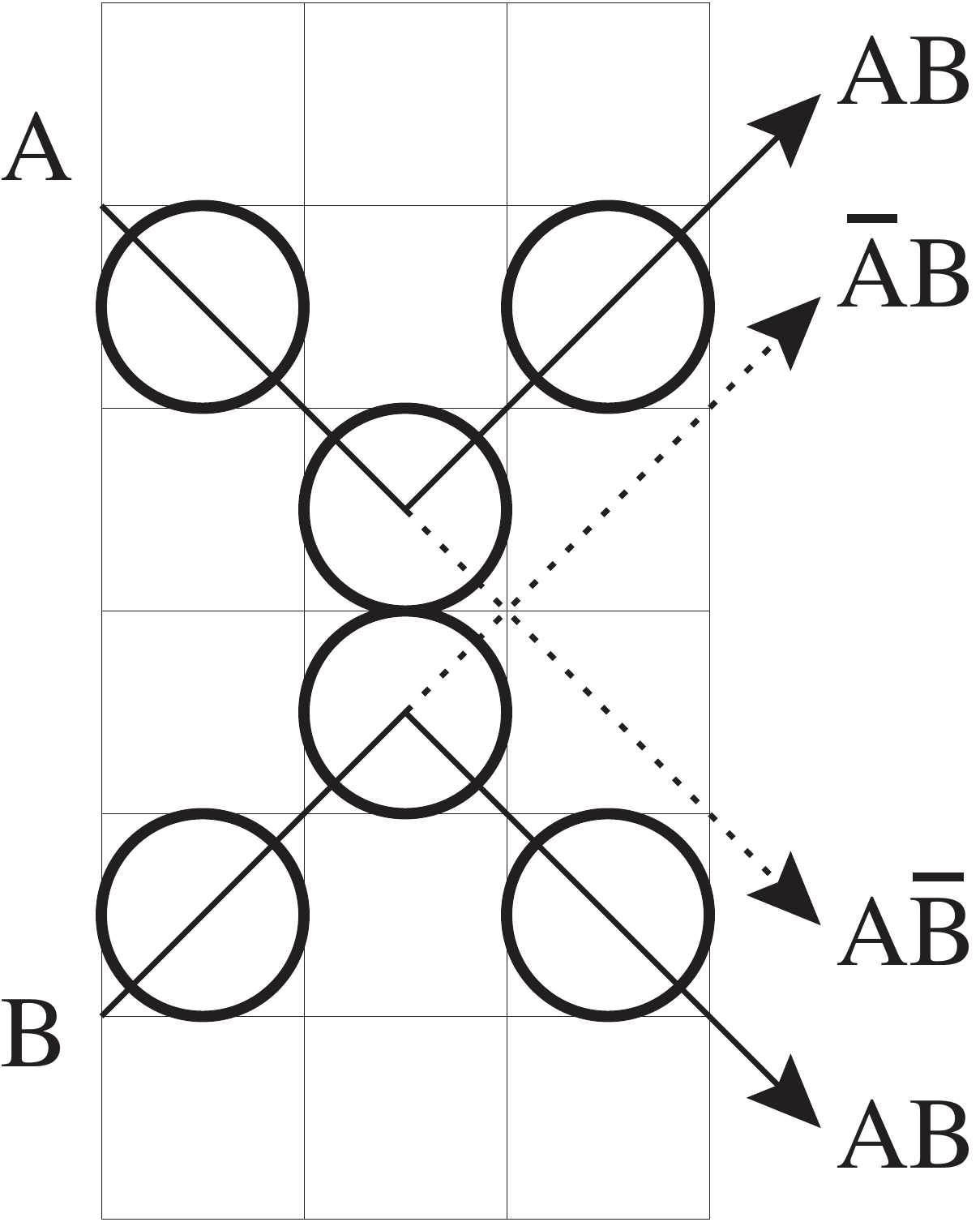} \hspace{.3in}\includegraphics[height=1.2in]{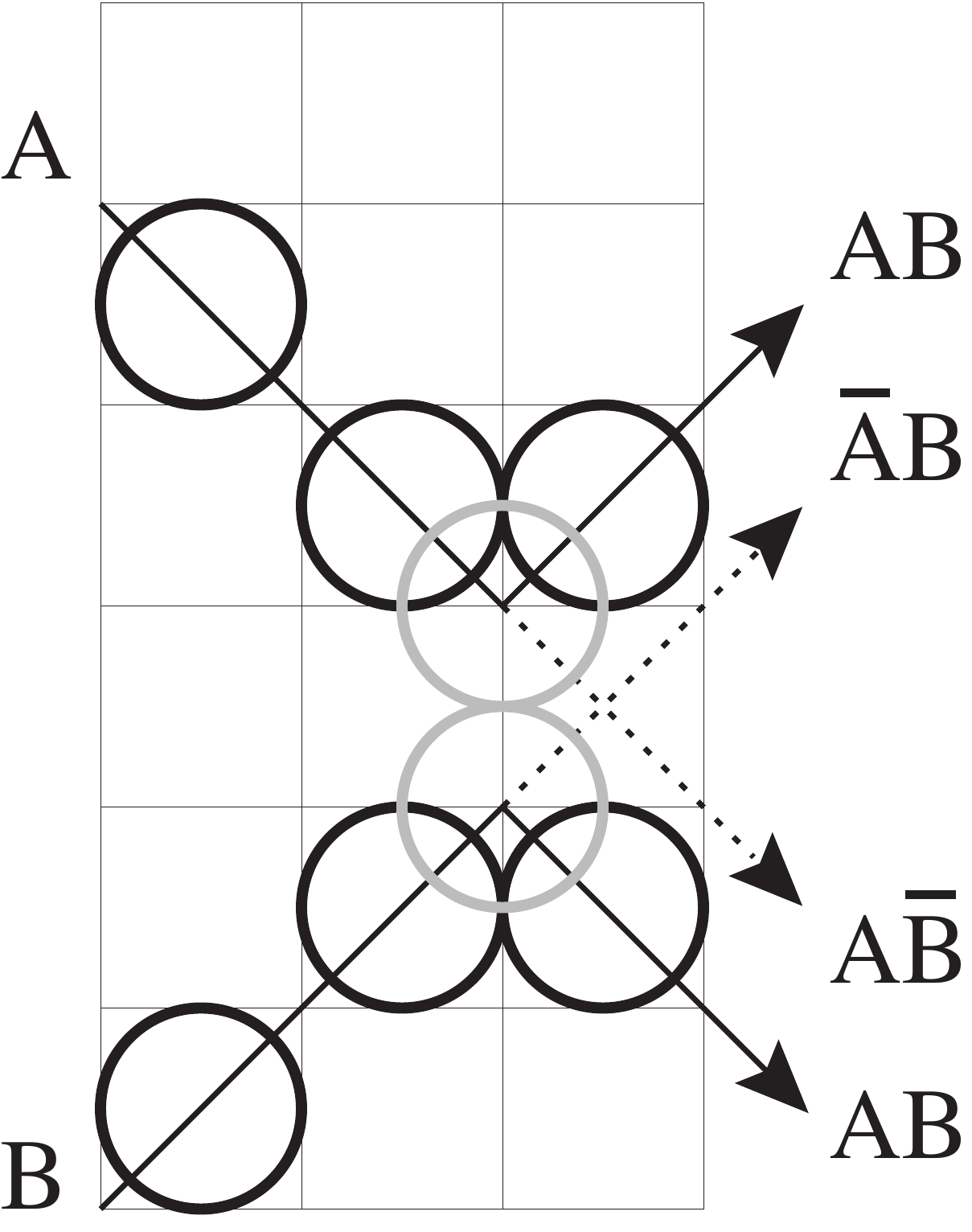}\hspace{.3in}\includegraphics[height=1.2in]{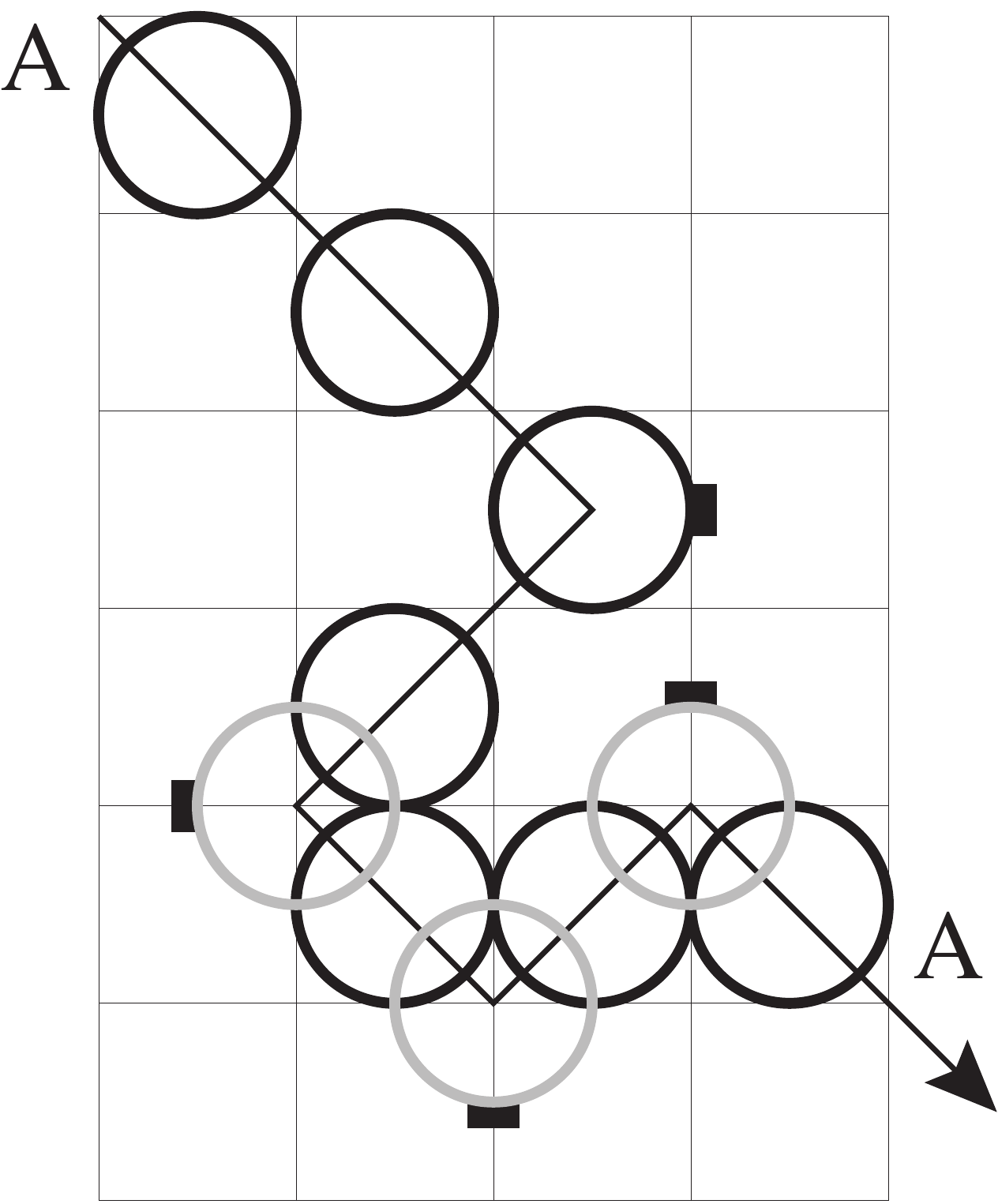}\hspace{.23in}\includegraphics[height=1.2in]{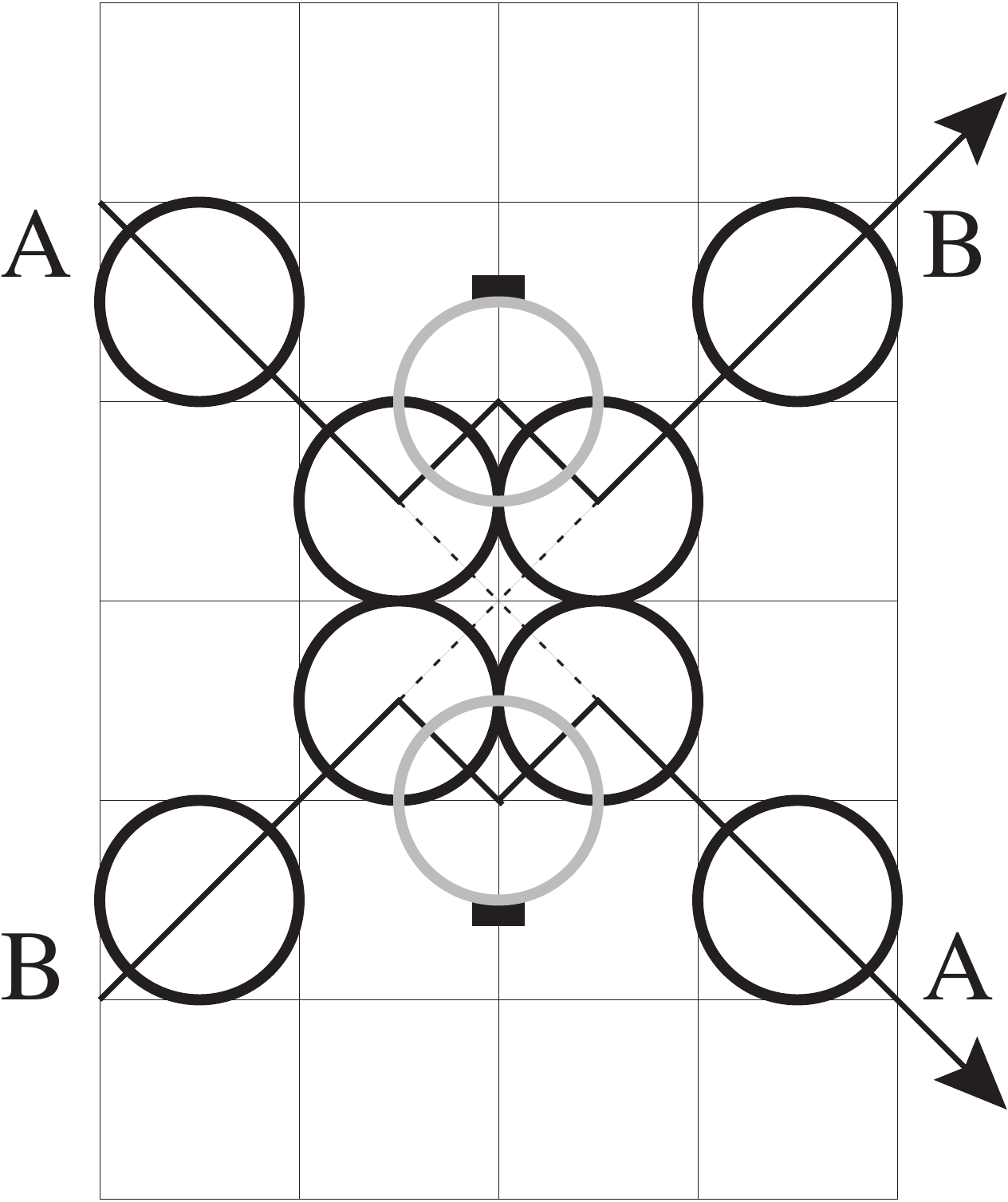} \end{minipage}}
  \caption{{\bf Fredkin's billiard ball model of computation.}  Perfect
    billiard balls moving on a lattice can perform reversible
    computation.  Ball presence is a 1, absence a 0.  Left to right, we
    show (a) a right-angle collision where different logical
    combinations come out at different places, (b) a logical collision
    that happens in between integer times, (c) infinitely massive
    mirrors used to route signals, and (d) mirrors used to allow signal
    paths to cross, regardless of the signal values.}
\label{fig.fredkin}
\end{figure}  

In Figure~\ref{fig.fredkin} we illustrate the lattice gas that Ed
Fredkin invented \cite{bbm} to try to silence skeptics who claimed
reversible computing was physically impossible.  This is a perfectly
good reversible, particle and energy conserving classical mechanical
model that uses the collisions of hard spheres to perform computation:
ball or no-ball is a one or a zero, and collisions separate different
logical cases into separate paths.  The model uses infinitely massive
mirrors, which are allowed in non-relativistic classical mechanics, to
route signals. The equivalence to a discrete lattice gas is incomplete,
though, since certain collisions (such as head on ones) will take the
balls off the lattice---it is not enough to merely start all balls at
lattice locations.  The model can be completed by simply mandating that
all problematic cases cause balls to pass through each other without
interacting.  This is a general feature of sampled classical dynamics:
it is typically necessary to add a form of {\em classical tunneling} to
the continuous dynamics, in order to maintain its digital character.
From the point of view of the mathematical machinery of classical
mechanics, there is really nothing wrong with doing this: it doesn't
impact invertibility, conservations, relativity, etc.

There is still a problem, though, with turning the billiard ball model
into a lattice gas that acts as a faithful sampling of a classical
mechanical dynamics.  Because of the hard collisions, the number of
locations that need to be updated as a group in order to ensure
invertibility and particle conservation is rather large: in
Figure~\ref{fig.fredkin} (b), when the ball coming in at {\bf B} is in
the second column and about to interact, the next value of the location
marked {\bf AB} at the top depends on the presence or absence of the
ball coming from {\bf B}.  To implement this as a lattice gas requires
lattice sites that hold many particles, or the use of rather large
partitions.  This structure also implies extra constraints, not present
in the continuous classical model, on the positions where collisions
can occur, so that particles only converge on places where all
particles can be updated as a group.

\subsubsection{Soft sphere model.}

\begin{figure}[t]{ 
\begin{minipage}{\textwidth}
  \includegraphics[height=1in]{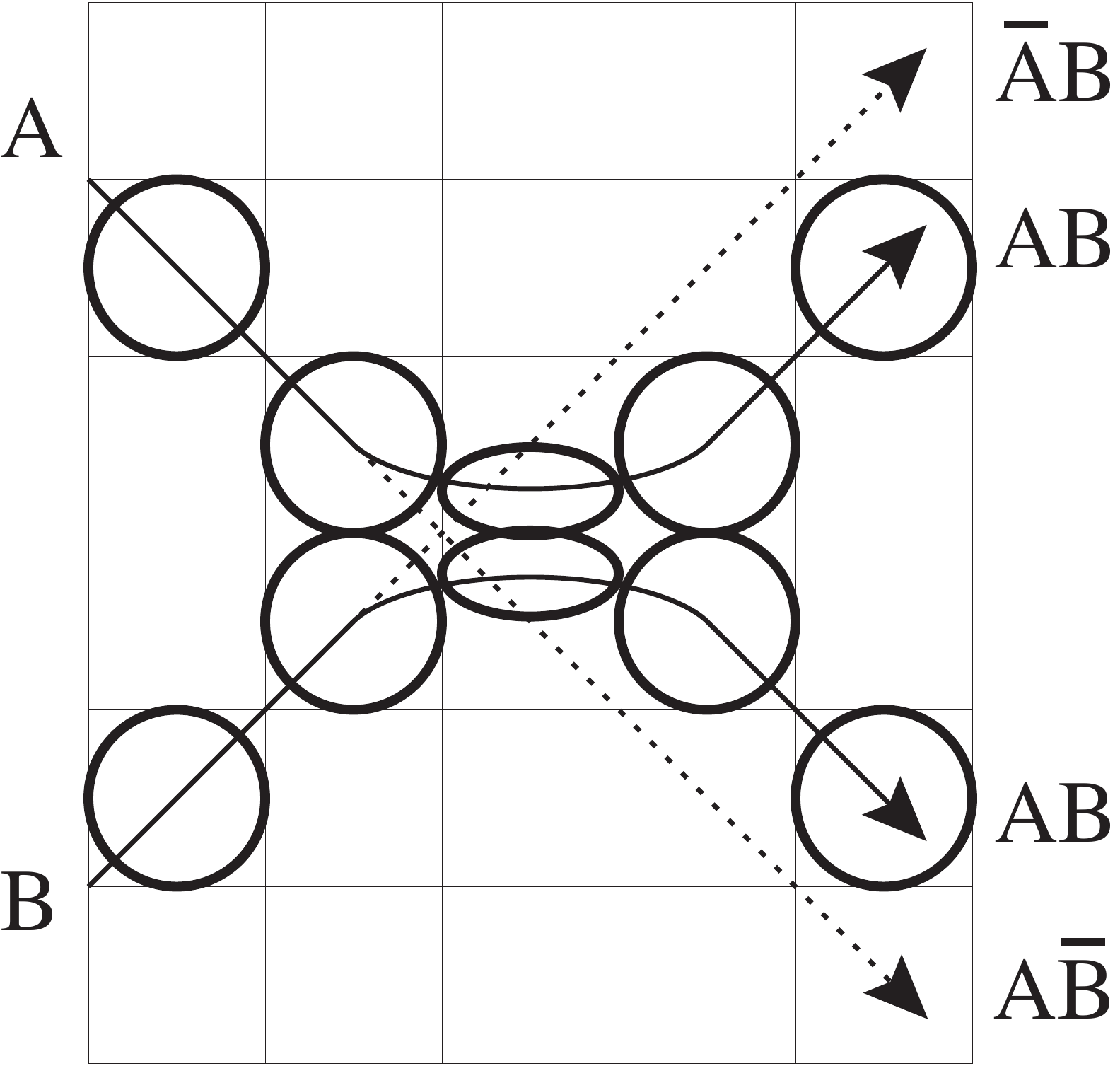} \hspace{.25in}
  \includegraphics[height=1in]{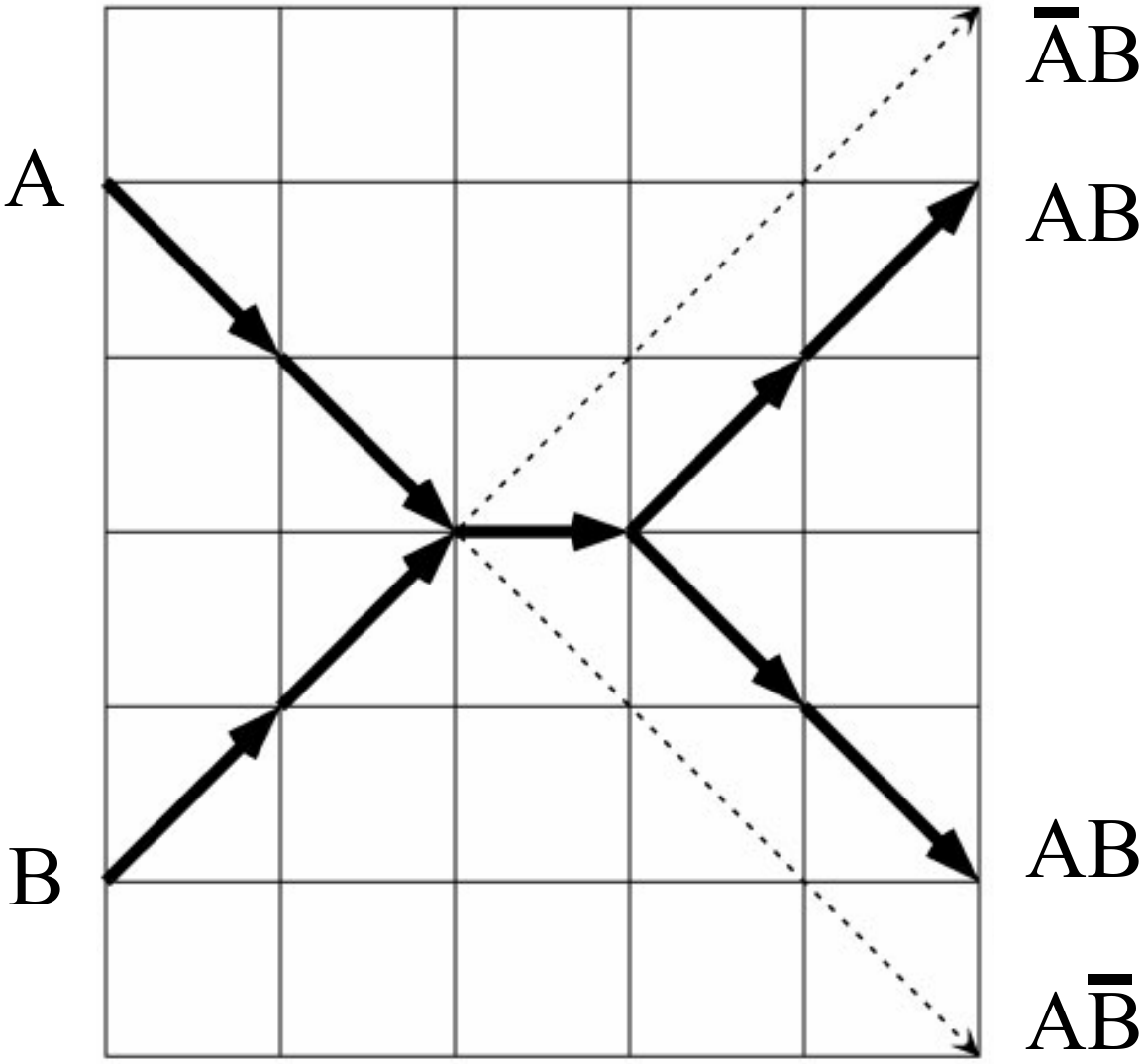}\hspace{.25in}
  \includegraphics[height=.7in]{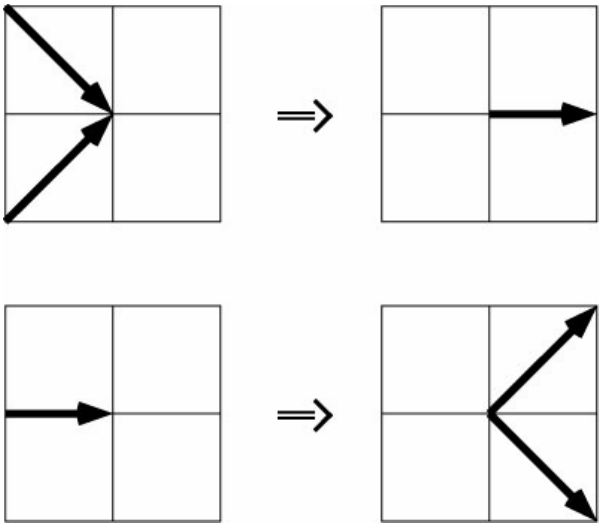}\hspace{.25in}
  \includegraphics[height=1in]{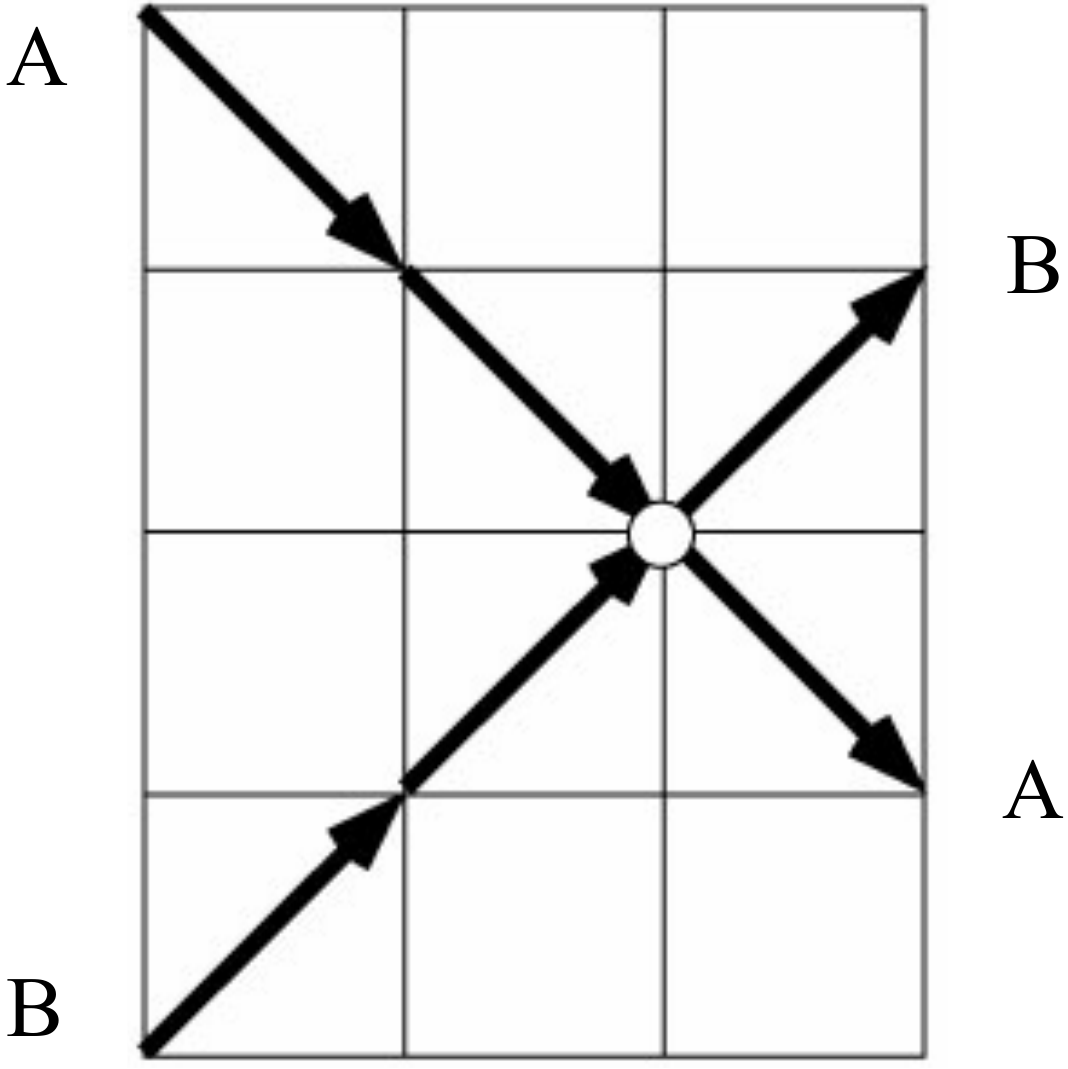} \end{minipage}}
  \caption{{\bf Soft sphere model of computation.}  Compressible balls
    collide.  Left to right we show (a) collisions displace the
    colliding balls inwards, putting the AB cases on different paths,
    (b) we recast this as a lattice gas, with particles located at the
    tails of the arrows, (c) there is interaction in only two cases
    (and their rotations); otherwise all particles move straight, (d)
    adding a rest particle to the model allows particle paths to cross;
    this still follows the rule ``otherwise all particles move
    straight.''}
\label{fig.ssm}
\end{figure}  

We can avoid all of these issues with a simple modification of the
classical billiard ball dynamics, illustrated in \ref{fig.ssm} (a).  If
we make the collisions very elastic, rather than hard, colliding
particles spend a finite amount of time colliding, and are deflected
onto inward paths, rather than outward as in the billiard ball model.
This soft sphere model \cite{ssm} is equivalent to a lattice gas where
interactions can happen at a point, as in Figure~\ref{fig.lga}
(left-top).  The discrete and continuous models can exactly coincide
everywhere at integer times.

Figure~\ref{fig.ssm} (b) shows a direct translation of (a) into a
lattice gas.  As in (a), we have two streams of particles coming in at
{\bf A} and {\bf B} and depict the state at an integer time, so we see
particles at each stage of the collision at different points in space.
In the lattice gas diagram, the particles are located at the tails of
the arrows, and the arrows indicate direction information carried by
the particles.  The rule (c) is very simple: diagonal particles
colliding at right angles turn into horizontal particles and vice
versa---plus $90^\circ$ rotations of these cases.  In all other cases,
the particles pass through each other unchanged.  In (d) we add an
unmoving {\em rest particle} to the model, so we can place it at any
signal crossing to prevent interaction: the rule is unchanged, since
moving straight is already the behavior in ``all other cases.''  This
allows the model to perform computation without the addition of
separate infinite-mass mirrors \cite{ssm}.  Similar computing lattice
gases can be defined on other lattices, in 2D or 3D.

We can analyze the minimum energy and momentum for a unitary quantum
implementation of this reversible classical dynamics.  Looking at
Figure~\ref{fig.ssm} (b), we count state change and direction of motion
in the middle of each arrow---during the time when the particle is
moving freely between lattice sites.  In this way we always see a
single isolated particle moving with a definite velocity, and can apply
\eqn{mom} directly to get the minimum momentum.  Taking the time
between lattice sites as our unit of time, each particle motion
constitutes one change per unit time.  The particles moving diagonally
are the fastest moving particles, so we take their speed $v=1$ to get a
minimum energy model---they each have ideal (minimum) momentum $p=1$.
From the geometry of the model, we see that the horizontal particle
must then be moving at speed $1/\sqrt{2}$, and there is again one
change in a unit of time as it moves, so its ideal momentum from
\eqn{mom} is $\sqrt{2}$.  This agrees with conservation of momentum,
since each of the two incoming particles has a horizontal component of
momentum of $1/\sqrt{2}$.  The horizontally moving particle is moving
slower than light, and so has a mass.  By energy conservation, since
the sum of the incoming energies is $2$, that must be the energy of the
horizontal particle.  Then $m=\sqrt{E^2-p^2}=\sqrt{2}$ is its mass.
This is a classical mechanical system with intrinsic energy, momentum
and mass ({\em cf.}  \cite{ideal-energy}).

\subsection{Sampled field}

\begin{figure}{
  \includegraphics[height=1.50in]{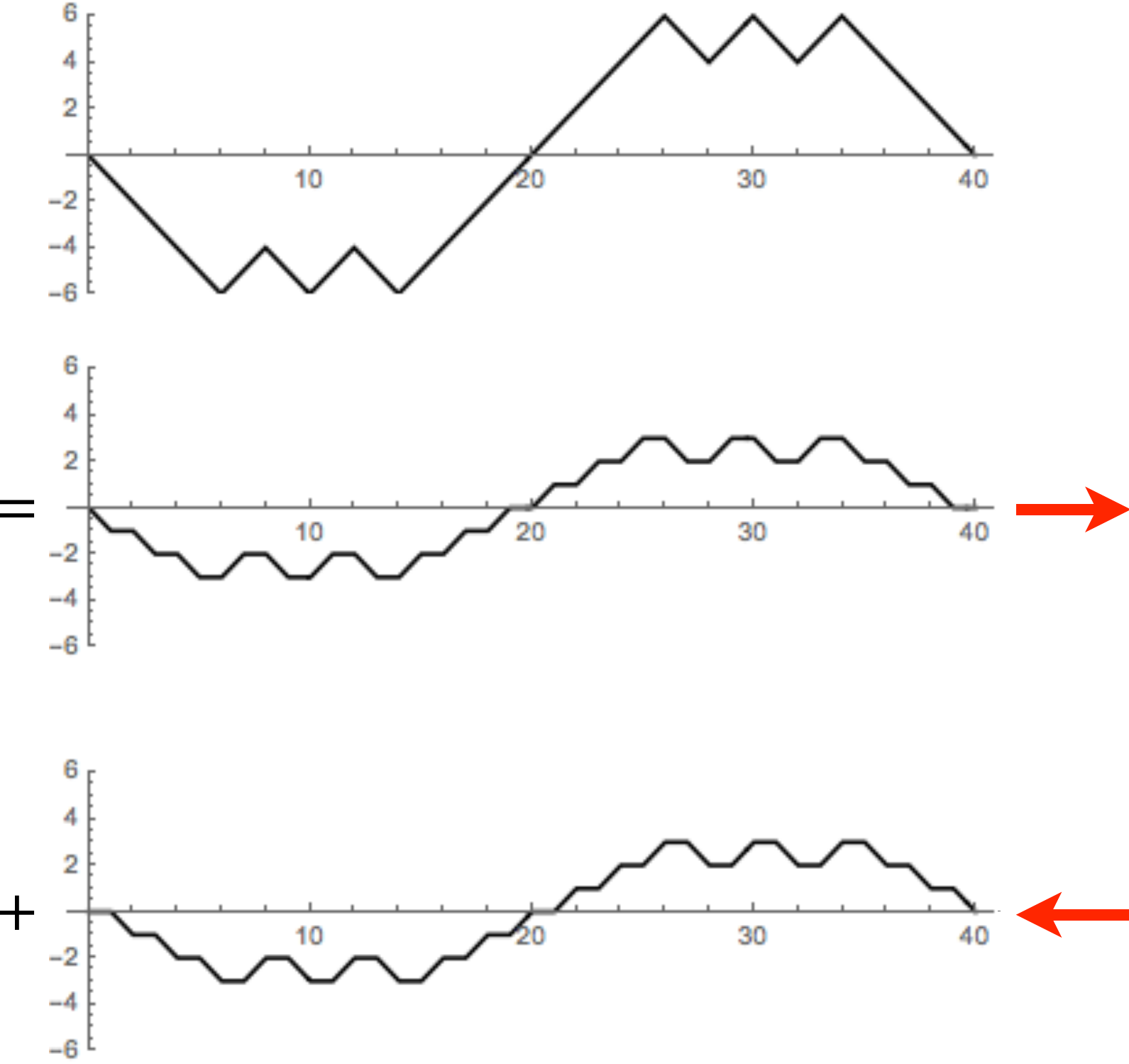} \hspace{.28in}
  \includegraphics[height=1.50in]{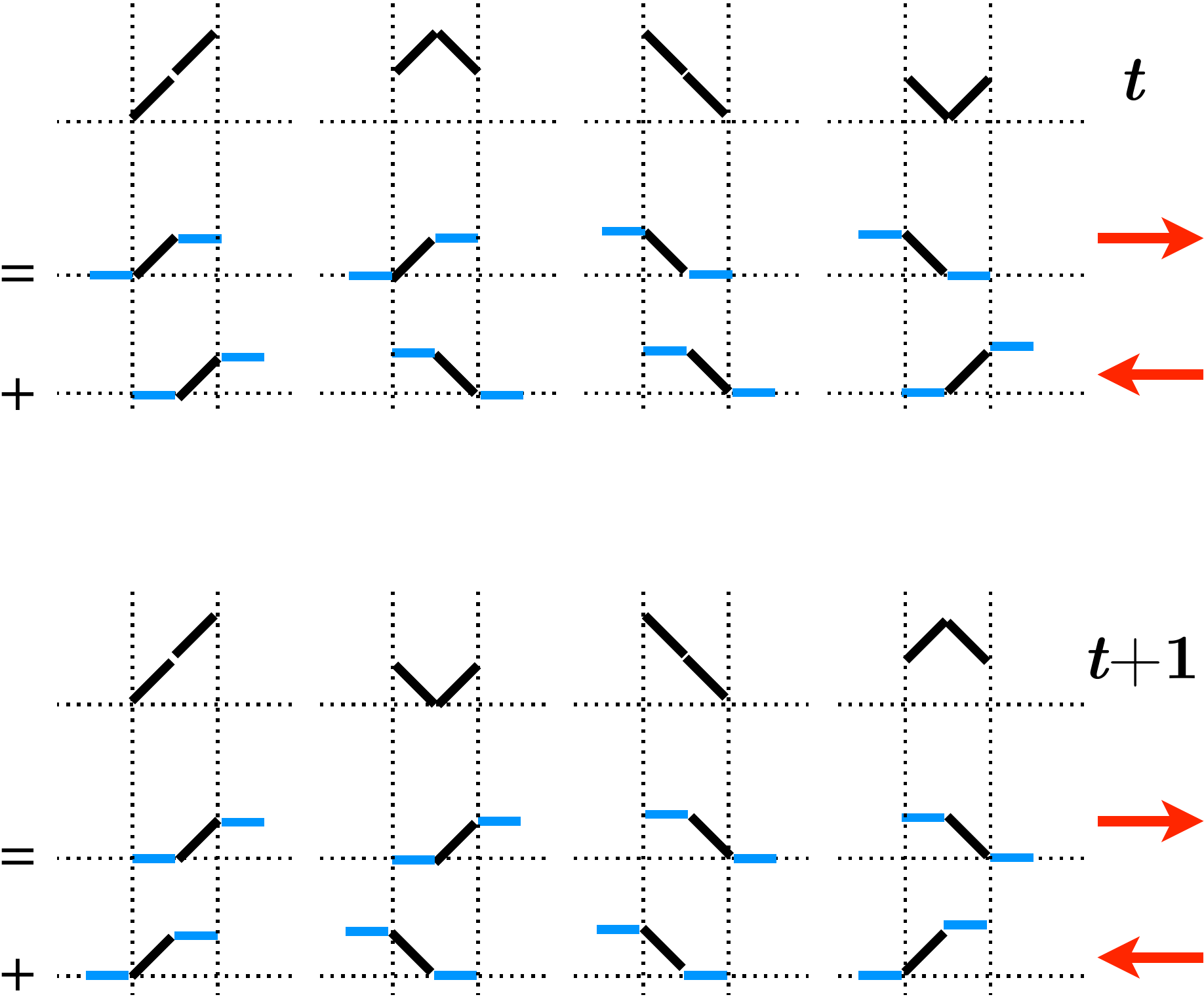}\hspace{.35in}
  \includegraphics[height=1.50in]{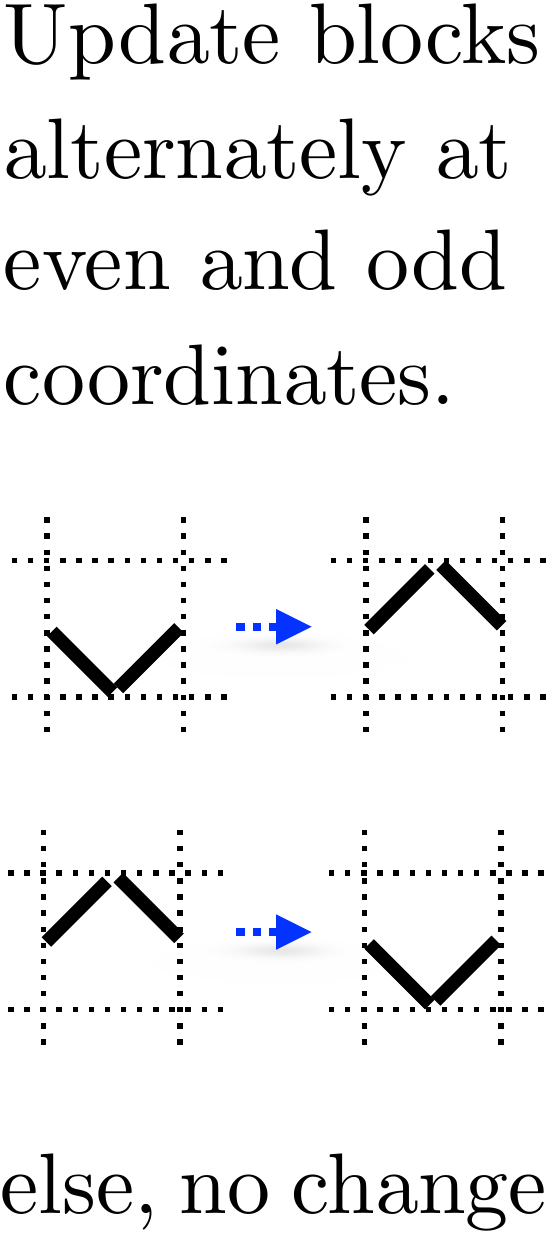}} \caption{{\bf Continuous
      wave dynamics equivalent to a finite-state dynamics.}  {\em
      Left:} Any one-dimensional wave obeying the continuous wave
    equation is a superposition of a rightgoing and a leftgoing wave.
    We constrain the two components to always give a discrete sum at
    integer times, so we can sample then.  {\em Middle:} Each component
    wave alternates flat intervals with intervals that have a slope of
    $\pm 1$.  There are only four cases possible at integer times.
    {\em Right:} Block rule for directly evolving the sum wave.}
\label{fig.wave}
\end{figure}  

We discuss here a simple classical field dynamics in which constraints
on the continuous initial state make it equivalent to a reversible
lattice gas at integer times and positions ({\em cf.}
\cite{ideal-energy,hh,toffoli-string,thesis}).  This example
illustrates the mechanism behind a phenomenon that was discovered
experimentally in reversible cellular automata models: the spontaneous
appearance of realistic waves \cite{cam-book,thesis,cc}.

In Figure~\ref{fig.wave} (left) we show, at the top, a continuous wave
that is the superposition of continuously shifting rightgoing and
leftgoing waves---we assume periodic boundaries so what shifts off one
edge reappears at the other.  It is a general property of the one
dimensional wave equation that any solution is a superposition of an
unchanging rightgoing waveform and an unchanging leftgoing waveform.
In this case, each of the two component waves contain segments that
have slope $0$, alternating with segments that have slope $+1$ or $-1$.
As the component waves shift continuously in space, at certain moments
the slope $0$'s of each wave align with the $\pm 1$'s of the other and
we get a discrete sum, composed only of segments with slope $\pm 1$.

Because of the discrete constraints, at integer moments of time the
space can be partitioned into pairs of adjacent columns where the
non-zero slopes are converging towards the center of each partition.
This is illustrated in Figure~\ref{fig.wave} (Middle).  At these times,
the next integer-time configuration for each pair of columns is
completely determined by the current configuration, and doesn't depend
on any information outside the pair.  We see, from
Figure~\ref{fig.wave} (Middle), that there are only four distinct
cases, and only two of them change the sum: slopes \verb|\/| turn into
\verb|/\| and vice versa.  We flip hills and troughs; nothing else
changes.

Figure~\ref{fig.wave} (Right) is thus the evolution rule for a {\em
  discrete string dynamics} that exactly follows the continuous wave
equation at integer times and positions.  The rightgoing and leftgoing
waves can be reconstructed from the sum---only the sum is evolved as a
lattice gas.  This {\em flip} rule must be applied alternately to the
two possible partitions into pairs of columns.  The rule works just as
well, though, if we also partition pairs of rows, so the rule is
separately applied to $2\times 2$ blocks.

\subsubsection{Transverse motion.}

It is interesting to analyze the energy and momentum for a discrete
string, evolving under the {\em flip} rule, that has a net motion up or
down.  If we decompose such a string into a superposition of a
rightgoing and a leftgoing wave, we find that each has a net slope
across the space.  As long as these two net slopes add to zero, the
string itself will have no net slope and so meets itself correctly at
the edges.  The rightgoing and leftgoing waves should be thought of as
infinite repeating patterns of slopes, rather than as meeting
themselves at the edges of one period of the string.

Let $2N$ be the width of one period of the string, in units of the
width of a slope segment.  The repeating pattern of the string can be
decomposed into the sum of a repeating pattern of $N$ non-zero
rightgoing slopes and $N$ non-zero leftgoing slopes.  Let $R_+$ be the
number of the $N$ rightgoing slopes that are positive, $R_-$ the number
that are negative, and similarly with $L_+$ and $L_-$ for the leftgoing
wave.  The net slope of the string will be zero as long as $R_+ + L_+=
R_- + L_-$.

Our unit of distance is the width ($=$ height) of a slope segment; our
unit of time the time needed for a slope segment to move one width.
Thus the rightgoing wave will be displaced upward a distance $R_- -R_+$
in time $2N$, and the leftgoing wave up by a distance $L_+ - L_-$ in
the same time, so the net upward displacement of the string will be
$D=R_- -R_+ +L_+ - L_-$ in time $2N$.

We can calculate this another way.  Consider a partition of the
rightgoing and leftgoing waves into pairs of columns at an integer
time.  A pair containing slopes \verb|\\| or \verb|//| contributes a
net of zero to $D$, since $R_+$ and $L_+$ enter with different signs,
as do $R_-$ and $L_-$.  Thus all of the net motion can be attributed to
the \verb|\/| and \verb|/\| cases, which respectively contribute $+2$
and $-2$. If there are $N_\up$ partitions ready to flip up, and $N_\dn$
partitions ready to flip down, the net upward velocity of the string is
\begin{equation}\label{eq.vnn}
  v = \frac{D}{2N} = \frac{N_\up - N_\dn}{N}\;.
\end{equation}

\subsubsection{String momentum and energy.}

Using \eqn{mom} we can assign minimum momenta.  We take the maximum of
$v$ to be the speed of light (assuming this string speed is physically
possible), and get one unit of momentum upward for each isolated motion
\verb|\/|$\to$\verb|/\|, one downward for each \verb|/\|$\to$\verb|\/|,
so net momentum upward is
\begin{equation}\label{eq.pnn}
  p = N_\up - N_\dn\;.
\end{equation}
Treating the overall motion of the string as that of a relativistic
particle, the total energy $E=p/v$, so from \eqn{vnn} and \eqn{pnn},
the total relativistic energy
\begin{equation}
E=N\;.
\end{equation}

We can compare this to the energy of the string, treated as a
continuous classical mechanical system.  Using the notation $\Psi_t$ to
denote partial derivative with respect to $t$, and with $c=1$, the
continuous 1D wave equation is $\Psi_{tt}=\Psi_{xx}$, and the classical
energy density of the string is proportional to $\Psi_{t}^2 +
\Psi_{x}^2$.  Since all of the slopes are $\pm 1$ the integral of this
sum of squares at integer times is just $4N$, twice the length of the
string.  The integral is also the same at all intermediate times.  This
follows from $\Psi_{tt} =\Psi_{xx}$ but we can also see this directly.
Let $l(x,t)$ be the amplitude of the leftgoing wave, and $r(x,t)$ the
rightgoing.  Then $\Psi=l+r$, $\Psi_x=l_x+r_x$ and $\Psi_t=l_x-r_x$, so
$\Psi_{t}^2 + \Psi_{x}^2 = 2(l_x^2 + r_x^2)$, and the integral is $4N$.

Similarly, the classical Lagrangian density ${\mathcal L}\propto
\Psi_{t}^2 - \Psi_{x}^2$ can be compared.  In this case, again using
$\Psi=l+r$ etc., ${\mathcal L} \propto l_x r_x$.  At integer times the
integral over the width of the space is zero, since every non-zero
slope is aligned with a zero slope.  Halfway between integer times,
half the columns contain pairs of non-zero slopes that are passing each
other, the other half contain pairs of zero slopes, which don't
contribute to the integral.  Each non-zero pair of equal slopes
contributes $+1$ to the integral; each pair of unequal slopes
contributes $-1$.  This is a count of potential minus kinetic energy:
halfway through a transition \verb|\/|$\to$\verb|/\| or
\verb|/\|$\to$\verb|\/| of a continuous string, all of the energy would
be kinetic, and for an unchanging length of stretched string \verb|//|
or \verb|\\| all energy would be potential.

\subsubsection{Rest frame energy.}

For a string moving according to this discrete wave dynamics, there is
a maximum amount of internal evolution of string configurations when
the net vertical velocity is zero, and no internal evolution when the
string is moving vertically as fast as possible, since then there is a
unique configuration
\begin{equation}
  \cdots \verb|\/\/\/\/\/\/\/\/\/\/\/\/\/| \cdots
\end{equation}
that moves upward at the speed of light.  The rest frame energy of
the string characterizes how much dynamics can happen internally.
This is given directly by $\sqrt{E^2-p^2}$ as
\begin{equation}\label{eq.er}
  E_r \; = \; \sqrt{N^2 - (N_\dn - N_\up)^2}\;.
\end{equation}
The more the string has a net motion, the less rest energy it has.
This makes the string a strange relativistic particle, since its total
energy is independent of its speed!  In the relation $E_r=E/\gamma$, we
have $E_r\to 0$ as $1/\gamma\to 0$.

This is not surprising, though, given the limited scope of the model:
it includes no interactions that could cause the string's travel to be
sped up or slowed down.  A realistic model of particle motion would
have to include interactions with other systems, that can add or remove
particle energy while conserving energy and momentum overall.  We might
still use the simple string model, though, as a proxy for a more
realistic finite-state model of inertia, by simply specifying a
statistical interaction that can change the length of the string to
change $E$.

\subsubsection{Statistical inertia.}

To contemplate a statistical coupling to another system, it is helpful
to recast the analysis of the string model in a population-statistics
format.  Recall that, for a given population of rightgoing and
leftgoing slopes, the number going each way is the same and, for the
string to meet itself at the edges, the sum of all the slopes is zero.
Together, these two constraints imply that $L_-=R_+$ and $L_+=R_-$: the
population statistics for rightgoing and leftgoing waves are mirror
images.  Therefore we can analyze the motion looking at the statistics
for just one of the component waves.  From \eqn{vnn}, using $N_\up -
N_\dn = D/2=L_+ - L_-$ and letting $p_+=L_+/N$ and $p_-=L_-/N$,
\begin{equation}
  v \; = \; p_+ - p_-\;,
\end{equation}
with $p_++p_-=1$.  Here $p_+$ is the average frequency of upward steps
per unit time, and $p_-$ the frequency of downward.  To see this for
$p_+=L_+/N$, notice that each leftgoing \verb|/| contributes one unit
of upward motion in the course of $2N$ steps of evolution, and so does
the mirror image \verb|\| moving rightward, so together they take us
one position up in $N$ steps; $p_-$ is similar.  Thus the transverse
motion of the string is like a one-dimensional random walk---which is
known to exhibit similarities to relativistic particle motion
\cite{curious-properties,smith,toffoli-string}.

These frequencies could become true probabilities if the populations
were statistically coupled to some environment.  For example, imagine
the string acting as a mass coupled to a spring, to form an oscillating
system.  As the spring stretches it slows down the mass, removing
energy and changing the bias $p_+ - p_-=v$.  Eventually it turns the
mass around and speeds it up, etc.  If the populations are stochastic
the velocity and energy determine the entropy of the string, which
would change cyclically with time in an oscillator.  This introduces a
rather thermodynamic flavor into a discussion of inertia in classical
mechanics.



\section{Conclusions}

Reversible lattice gas dynamics derived by sampling classical
mechanical evolution are foundational models for all of mechanics, in
the same way that classical lattice gases have long been foundational
for statistical mechanics.  They are pardoxically both continuous and
discrete, both classical and quantum.  They have an intrinsic energy
and momentum based on counting classical state change given by the
minimum allowed by the general properties of energy and momentum in
quantum mechanics.  They are non-trivial models and can in fact be
computation universal.

These models are foundational rather than fundamental.  We are not at
all suggesting that nature is, at base, a classical cellular automaton
\cite{dm,thooft,nks}, but rather that these reversible classical
systems that are also special cases of unitary quantum systems provide
a simplified context in which to study the foundations of mechanics.
This is exactly the role that classical lattice gases play in
statistical mechanics: classical special cases of quantum systems
\cite[\S 2.4]{ruelle}.  These dynamical counterparts should be studied
not just for academic or pedagogical interest---though that is reason
enough---but also because there are fundamental informational issues
that are not understood in physics, and these models introduce a
realistic counting of distinct quantum states into classical mechanics
and classical field theory.  It might be, for example, that some of the
(very macroscopic) informational paradoxes of general relativity depend
only on the reversibility of the dynamics, and not on the full
unitarity of quantum evolution.

The present analysis only scratches the surface of what seems to be a
rich field.  Even discovering what kinds of macroscopic phenomena {\em
  cannot} be modeled in this manner may tell us something about the
essential role that quantum mechanics plays, that couldn't be played by
a classical informational substratum.  These finite-state classical
mechanical models also turn some foundational questions on their head,
since they can be regarded as special cases of unitary quantum
evolution, rather than macroscopic decoherent limits.

\subsubsection*{Acknowledgments.}  I thank Ed Fredkin and Tom
Toffoli for pioneering and inspiring these ideas, and Gerald Sussman
for many wonderful discussions.

\end{document}